\documentclass[aps,prb,amsmath,amssymb,superscriptaddress,preprint]{revtex4-1}

\usepackage{amssymb}
\usepackage{yfonts}
\usepackage{graphicx}
\usepackage{amsmath}
\usepackage{color}
\usepackage{epstopdf}
\usepackage{ulem}

\usepackage{bm}% bold math
\usepackage{appendix}
\begin{document}

\title{Spin Wave Generation via Localized Spin-Orbit Torque in an Antiferromagnet-Topological Insulator Heterostructure}

\author{Xinyi Xu}
\affiliation{Department of Electrical and Computer Engineering, North Carolina State University, Raleigh, NC 27695, USA}

\author{Yuriy G. Semenov}
\affiliation{Department of Electrical and Computer Engineering, North Carolina State University, Raleigh, NC 27695, USA}
\affiliation{V. Lashkaryov Institute of Semiconductor Physics,  National Academy of Sciences of Ukraine, Kyiv 03680, Ukraine}

\author{Ki Wook Kim}\email{kwk@ncsu.edu}
\affiliation{Department of Electrical and Computer Engineering, North Carolina State University, Raleigh, NC 27695, USA}
\affiliation{Department of Physics, North Carolina State University, Raleigh, NC 27695, USA}

\begin{abstract}
The spin-orbit torque induced by a topological insulator (TI) is theoretically examined for spin wave generation in a neighboring antiferromagnetic thin film. The investigation is based on the micromagnetic simulation of N\'{e}el vector dynamics and the analysis of transport properties in the TI. The results clearly illustrate that propagating spin waves can be achieved in the antiferromagnetic thin-film strip through localized excitation, traveling over a long distance.  The oscillation amplitude gradually decays due to the non-zero damping as the N\'{e}el vector precesses around the magnetic easy axis with a fixed frequency. The frequency is also found to be tunable via the strength of the driving electrical current density.  While both the bulk and the surface states of the TI contribute to induce the effective torque, the calculation indicates that the surface current plays a dominant role over the bulk counterpart except in the heavily degenerate cases.  Compared to the more commonly applied heavy metals, the use of a TI  can substantially reduce the threshold current density to overcome the magnetic anisotropy, making it an efficient choice for spin wave generation.  The N\'{e}el vector dynamics in the nano-oscillator geometry are examined as well.
\end{abstract}
\maketitle

\section{Introduction}
Spin waves have recently attracted much attention as a potential carrier of information with low energy dissipation.  They can transport a pure spin current without involving the charge flow along with the possibility to manipulate the amplitude and phase.~\cite{Chumak2015,Kruglyak} The excitation of spin waves has been achieved via the optical, thermal or electrical methods.~\cite{Lenk,Seki,Cherepov,Jamali}  The mechanisms that take advantages of the electrically induced effective torque offer one of the most efficient approaches without introducing bulky external antenna.~\cite{Demidov2016,Xu2019}  In a heterostructure consisting of a magnetic layer and a strongly spin-orbit coupled (SOC) material,  the spin-orbit torque (SOT) resulting from the spin dependent electron trajectories can manipulate the magnetic state of the magnet.~\cite{Demidov2016}  For instance, auto-oscillations can be driven via the SOT, resulting in the propagating spin waves as it has been demonstrated experimentally in a ferromagnetic structure.~\cite{Fulara2019}  The spin wave frequency, although limited by the characteristic properties of the waveguide, is proportional to the strength of the SOT. Evidently,  a SOC layer that can realize a stronger torque for a given driving current density is highly desirable for a range of applications with an obvious ramification on the effectiveness of electrical control.

Heavy metals such as Pt and W are the commonly used SOC materials with the spin-Hall angle $\theta_\mathrm{SH}$ typically in the range of $0.012 \sim 0.12$,~\cite{Wang2014} requiring a relatively high driving current density ($\sim 10^7$~A/cm$^2$)  to overcome the magnetic anisotropy.  In comparison,  recent investigations have indicated that a topological insulator (TI) can induce the SOT with a higher efficiency than the heavy metals $-$ an order of magnitude higher or even  more  at room temperature.~\cite{Han,Wu,Yang17}  The experimentally determined values of $\theta_\mathrm{SH} > 1 $  have been reported in the literature,~\cite{Mellnik,Dc} making the TI a promising alternative to the more conventional heavy metals.~\cite{Band}
It is well known that the TIs are characterized by the insulating bulk and two-dimensional (2D) semi-metallic surface states.  In particular,  the topologically protected surface electronic states with a linear dispersion exhibit a strong magnetoelectric effect via the inherent spin-momentum interlock.~\cite{Moore}  While the bulk states are frequently overlooked, the bias-driven in-plane current tends to flow both in the bulk and on the surface of the TI since the energy separation between them is relatively small.  In fact, the electrical current in the bulk can induce the spin torque in a manner similar to that in a heavy metal (i.e., the spin-Hall effect).~\cite{Ghosh18,Ghosh19}
Evidently, the bulk contribution depends on the details of the band structure including the band separation and the position of the Fermi level. As the Fermi level can be modulated externally, their impact on  the induced SOT can show a range of responses, necessitating a careful consideration.

In this work, we theoretically explore the generation of traveling spin waves in a magnetic strip by exploiting the  localized excitation through a TI layer in place of a heavy metal.   Both the bulk and surface states are examined as the source of the SOT for a range of parameters such as the damping constant and the Fermi level position.  The thin-film waveguide structure based on an  A-type antiferromanget (AFM) with uniaxial easy-axis anisotropy is considered as the primary example, while the analysis can be readily extended to the ferromangets (FMs).  Compare to the FM counterparts, the AFMs are associated with higher resonance frequencies up to THz, faster switching speeds, and higher energy efficiencies.~\cite{Jungwirth2016}  Furthermore, they are not subject to the demagnetization field, offering a promising medium for spin wave propagation.   The investigation takes advantage of the micromagnetic simulations along with the analysis on the TI properties.   The results clearly illustrate the possibility of efficient spin wave generation that can travel over a long distance as well as its efficient control via the TI induced SOT.   The role of the bulk vs.\ surface currents is also elucidated which appears to be consistent with a recent experimental report.~\cite{Wu}

\section{Theoretical Model}
The dynamics of magnetization can be described by using the Landau-Lifshitz-Gilbert (LLG) equation as:
\begin{equation}
\frac{\partial \mathbf{m}}{\partial t} = - \gamma\mathbf{m}\times \mathbf{H}_\mathrm{eff}+\alpha\mathbf{m}\times\frac{\partial \mathbf{m}}{\partial t}+\mathbf{T}_\mathrm{SOC} .  \label{LLG}
\end{equation}
Here, $\mathbf{m}$ denotes the reduced or normalized local magnetization vector,  $\gamma$ is the gyromagnetic ratio, $\alpha$ refers to the Gilbert damping constant, and the macroscopic effective field $\mathbf{H}_\mathrm{eff}\propto\frac{\partial H}{\partial \mathbf{m}}$ is obtained from the Hamiltonian $H$ of the considered system that accounts for the exchange interaction and the anisotropy energy.  In the case of an AFM, sublattices $A$ and $B$ are antiferromagnetically coupled to each other via the exchange interaction, giving the normalized N\'{e}el vector $\mathbf{n}$ as $\mathbf{m}_{A} - \mathbf{m}_{B}$. The last term $\textbf{T}_\mathrm{SOC}$ in Eq.~({\ref{LLG}}) corresponds to the  SOT induced by an SOC material (in this case, a TI) that can be further separated into the anti-damping torque $\mathbf{T}_\mathrm{ad}$ and the field-like torque $\mathbf{T}_\mathrm{fl}$ depending on the actual physical processes (i.e., $\textbf{T}_\mathrm{SOC} = \mathbf{T}_\mathrm{ad} + \mathbf{T}_\mathrm{fl}$).

A schematic of the AFM/TI heterostructure under consideration is shown in Fig.~1.  As depicted,  an electrical current can flow in both the TI bulk and surface along the $ {x}$ axis when an appropriate bias voltage is applied.  The driving current is assumed to be direct (i.e.,\textit{ dc}) in comparison to the approach with a radio frequency source explored in the recent studies.~\cite{Wang19}
The induced SOT can excite the localized oscillations of the N\'{e}el vector in the AFM region directly in contact with the TI, which may subsequently be guided along the AFM strip with the easy axis aligned in the same direction (i.e., $y$).  To ensure the uniformity in the excited waves, the thickness of the magnetic film in the $z$ direction needs to be sufficiently smaller than the typical AFM exchange length of tens of nanometers.~\cite{Abo}  An insulating or dielectric material is preferred for the AFM strip to avoid the current shunting (thus, loss) in the excitation region.  Two key mechanisms that can lead to the desired N\'{e}el vector oscillations are  the spin-Hall effect in the TI bulk and  the proximity effect driven by the coupling with the magnet on the surface.~\cite{Ghosh18,Semenov2014}

The spin-Hall effect taking place in the bulk generates a spin current in the $z$ direction.  Subsequent injection into the adjacent magnetic layer leads to the N\'{e}el vector dynamics by exerting an anti-damping torque in the following form~\cite{Liu2011}
\begin{equation}
\mathbf{T}_\mathrm{ad}^\mathrm{B}= \gamma\frac{\hbar}{2q\mu_0 M_sd}\theta_\mathrm{SH} {J}^\mathrm{B} [\mathbf{m}\times (\hat{\mathbf{\sigma}} \times \mathbf{m})] ,  \label{Tb}
\end{equation}
where $q$ ($> 0$) is the electron charge, $\mu_0$ the vacuum permeability, $M_s$ the sublattice magnetization, $d$ the thickness of the AFM layer, ${J}^\mathrm{B}$ the bulk current density, and $\hat{\mathbf{\sigma}}$ the unit vector of the spin current polarization ($ \| \pm \hat{\mathbf{y}}$ when the electrons flow along the $\mp x$ direction). The product of spin-Hall angle  $\theta_\mathrm{SH}$ and  ${J}^\mathrm{B}$ gives the spin current density. The field-like torque that can also arise from the TI bulk current is neglected for its relatively minor contribution to the desired magnetization rotation or oscillation.  The spin-Hall angle $\theta_\mathrm{SH}$ of $0.15$  is chosen in Eq.~(\ref{Tb}) at room temperature, which is comparable to an upper-end value for the heavy metals.~\cite{Mellnik,Wu}  Note that this choice for $\theta_\mathrm{SH}$ (i.e., $0.15$) is to account for only the bulk effect.  An effective value extracted for the entire TI can be much larger as discussed briefly in the introduction.  The bulk contribution is included in the calculations unless mentioned explicitly otherwise. %which is slightly higher than heavy metal range 0.012∼0.12.

Concerning the effect of the TI surface, it is likely that multiple mechanisms of differing microscopic origins~\cite{Ghosh18,Duan2015,Mahfouzi} manifest simultaneously through the self-consistent interaction with the magnet.  As  their comprehensive account is yet to be achieved,  we adopt an empirical treatment based on an experimentally observed phenomenon $-$ the proximity induced anomalous Hall effect.~\cite{Duan2015,Zhu} The resulting $y$-directional Hall current ($ {J}_y$) introduces an additional spin polarization component
that can lead to the anti-damping behavior.    A simple expression is used to phenomenologically describe $ {J}_y$ in terms of the $x$-directional driving surface current ${J}^\mathrm{S}$;  i.e., $ {J}_y=-\beta_z m_z {J}^\mathrm{S}$.~\cite{Duan2015}
Here, $m_z$ signifies the ${z}$ component of AFM sublattice magnetization at the interface with the TI (which is non-zero in the A-type) and  $\beta_z$ is the ratio between the two current components that can be determined experimentally ($ \approx 0.06$).~\cite{Alegria,Chang,Zhang}  Given the empirical nature of the treatment, it may be possible to adjust this parameter $\beta_z$ to reflect the influence of other surface driven anti-damping mechanisms.   While the anomalous Hall effect is deemed absent in a collinear AFM due to the symmetry, the  A-type AFMs with ferromagnetic intra-plane coupling can actually have the magnetization on a surface essentially analogous to the FMs (i.e., non-zero).  Accordingly, the proximity interaction at the interface in the present case can be described as in the FM/TI bilayer structure studied earlier.~\cite{Ghosh18,Duan2015,Mahfouzi}
Combined with the field-like contribution, the total SOT induced by the surface states can then be written as~\cite{Duan2015}
\begin{equation}
\mathbf{T}^\mathrm{S}= \mathbf{T}_\mathrm{ad}^\mathrm{S} + \mathbf{T}_\mathrm{fl}^\mathrm{S}= -\frac{ \gamma G}{q\mu_0 M_s d v_F} ~ \delta t {J}^\mathrm{S}  ~\mathbf{ m} \times (\beta_z m_z\hat{\mathbf{ x}}+\hat{\mathbf{y}}) ,  \label{Tsur}
\end{equation}
where \textit{G} is the TI/magnet exchange coupling energy  and $v_F$ the Fermi velocity of the TI surface states.  The angular dependence in the anti-damping term is apparent from the expression, where the effective torque (i.e., $\mathbf{T}_\mathrm{ad}^\mathrm{S}$) becomes zero when  $m_z = 0$.  A similar dependence was reported in the literature.~\cite{Ndiaye}   Higher order terms such as those discussed in Ref.~[\onlinecite{Belashchenko}] are not considered.   Note that ${J}^\mathrm{S}$  in Eq.~(\ref{Tsur}) is modified to take a 3D equivalent form for direct comparison with ${J}^\mathrm{B}$.  More specifically, the actual surface current density (given by per unit length) is divided by the thickness $\delta t$ of the TI layer to convert it to a per-unit-area quantity (i.e., a 3D current density).~\cite{Wu}

%%%%%%%%%% spin wave results
The desired N\'{e}el vector dynamics are analyzed by numerically solving the LLG equation based on Object Oriented MicroMagnetic Framework (OOMMF).~\cite{oommf}   The AFM thin-film nanostrip is assumed to have the dimensions of $3\times 600 \times 1$ nm$^3$ with a perfectly absorbing boundary at both ends of the strip (i.e., no reflection to avoid the unnecessary complication), of which the first $90$~nm at the left end is excited by the SOT through the interaction with a TI layer in contact.  The easy $y$-axis anisotropy of $K_y = 20$ kJ/m$^3$ is adopted along with the exchange stiffness $A_\mathrm{ex}=5$ pJ/m, $M_s=350$ kA/m, and $\alpha=0.001$.  These values are within the range for a dielectric AFM such as NiO.

As for the TI, a relatively thick film of $10$~nm ($= \delta t$) is considered to ensure decoupling between top and bottom surfaces.  The Fermi velocity $v_F$ is set at $4.5\times 10^7$ cm/s for the surface states and the separation between the bulk conduction-band minimum $\varepsilon_C$ and the Dirac point $\varepsilon_D$ at $0.2$~eV, both of which are typical for well-known TIs such as Bi$_2$Se$_3$.  A parabolic energy dispersion is used for the bulk conduction band with an effective mass $m^*=0.15m_e$, where $m_e$ denotes the electron rest mass.~\cite{Butch}  The exchange coupling strength $G$ with the neighboring AFM is taken to be $40$ meV. Transport properties in the TI layer are evaluated by adopting a simple ohmic relation ${J}=q \mu_n n {E}$ with the mobilities  $\mu_n^S \approx \mu_n^B \approx 10^3$ cm$^2$/V$\cdot$s at room temperature.~\cite{Oh} While the electric field $E$ can clearly be modulated by the external bias as well,  a constant value of $12$~kV/cm is assumed to avoid proliferation of adjustable variables.  Accordingly, ${J}^\mathrm{B}$ and ${J}^\mathrm{S}$ are determined by the carrier densities, which can in turn be specified as a function of a single parameter, i.e., the Fermi level position $\varepsilon_F$.

\section{Results and Discussion}
Details of the excited spin wave dynamics and properties are illustrated in Fig.~2. With the application of a sufficiently large SOT at $t = 0$, the spins are driven away from the easy $y$-axis ($\mathbf{n}$ set initially in the $+y$ direction) to oscillate around it  in the excitation region, which then propagates along the channel. The precession angle does not reach 90$^\circ$ (i.e., $|n_y| > 0$) with non-zero damping ($\alpha \neq 0$) as the growth in the axial tilt via the anti-damping spin torque  is compensated by
the interaction with the magnetic moments in the unexcited part of the AFM strip.~\cite{Xu2019}  The precession can be around either the $+y$ [Figs.~2(a,d)] or $-y$ axis [which is preceded by the N\'{e}el vector flip; Figs.~2(b,e)] depending on the polarity of the excitation current (thus, $\hat{\mathbf{\sigma}}$).

Figures~2(a) and 2(b) show the N\'{e}el vector state obtained along the AFM strip at $t=0.1$~ns, by which time a steady spin wave is established.  The Fermi level $\varepsilon_F$ is assumed to be at $30$ meV above $\varepsilon_C$, which corresponds to the excitation current density (magnitude) of  ${J}^\mathrm{B}=5.3\times10^6$ A/cm$^2$ and  ${J}^\mathrm{S}=4.1\times 10^7$ A/cm$^2$.  The observed decay in the magnitude of  $n_z$  illustrates the continued reduction in the angle of rotation around the $y$ axis due to non-zero damping. Nevertheless, the traveling wave maintains constant wavelength and frequency.  The 3D illustrations of the N\'{e}el vector trajectories in the excitation region are plotted in Figs.~2(d) and 2(e), which correspond to the above mentioned cases of  $\hat{\mathbf{\sigma}} = + \hat{\mathbf{ y}} $ and $ - \hat{\mathbf{ y}} $, respectively.
The calculated dispersion relation between the angular frequency $\omega$ (=$2\pi f$) and the wavevector $k$ (=$2\pi/\lambda$) is essentially linear as indicated in Fig.~2(c), whose slope amounts to the magnon velocity $v_m$ ($=1.4\times10^6$ cm/s). The velocity $v_m$ is given by the characteristics of the AFM material as $\gamma\sqrt{H_\mathrm{ex}A_\mathrm{ex}/M_s}$, where the exchange field $H_\mathrm{ex}$ can be expressed further in terms of $A_\mathrm{ex}$ and $M_s$.~\cite{Gomonay} A larger current density (thus, the SOT) leads to a higher oscillation frequency of the N\'{e}el vector or the spin wave as well as a shorter wavelength.~\cite{Collet}

It is interesting to systematically examine the role of the often neglected bulk contribution to the SOT.  Figure 3(b) shows $J^S$ (line 1) and $J^B$ (line 2) calculated as a function of the Fermi level position with respect to the bulk conduction band minimum (i.e., $\Delta\varepsilon = \varepsilon_F - \varepsilon_C$). When $\Delta\varepsilon$ is negative (thus, $\varepsilon_F$ in the bulk bandgap), $J^B$ becomes very small and is thus not considered.  As $\varepsilon_F$ moves above $\varepsilon_C$, the current flow in the bulk goes up faster than the surface counterpart resulting in the gradual increase of the ratio $J^B/J^S$ (line 3).  Since the Fermi level in a  TI is typically chosen to be not far from the Dirac point $\varepsilon_D $, the surface current density is expected to be generally more pronounced than the bulk contribution in a sufficiently thin structure. Note that this observation is based on the assumption of $\varepsilon_C-\varepsilon_D=0.2$ eV.  A smaller separation between the bulk and surface bands can enhance the significance of $J^B$.

Nevertheless, the addition of even a relatively small bulk current may considerably alter the effectiveness of a TI as a SOC material.  For a more definitive understanding, a comparative analysis of the SOTs generated by the two mechanisms (i.e., the spin-Hall and anomalous Hall effects) is desired.  However, a direct term-by-term comparison is challenging due to the differences in the functional dependence.  When such details are ignored, Eqs.~(\ref{Tb}) and (\ref{Tsur}) suggest that the ratio between the bulk and surface induced anti-damping torques scales roughly as
\begin{equation}
|\frac{\mathbf{T}_\mathrm{ad}^{B}}{\mathbf{T}_\mathrm{ad}^{S}}| \approx  \frac{\hbar v_F}{2G} \frac{ \theta_\mathrm{SH}}{ \beta_z\delta t} |\frac{{J}^\mathrm {B}}{{J}^\mathrm {S}}| .
\end{equation}
The prefactor in front of ${{J}^\mathrm {B}}/{{J}^\mathrm {S}}$  is estimated to be around 1 with the numerical values discussed earlier.  Accordingly,  the Fermi level position [via Fig.~3(b)] appears to provide at least an approximate indicator for the contribution of the bulk states  although the actual impact on the N\'{e}el vector dynamics is determined by the physical details of the SOT processes.

For a quantitative evaluation of the macroscopic response, the frequencies of the excited spin waves are compared as a function of the Fermi level position $\Delta\varepsilon$ by using the micromagnetic simulations.  Figure~3(c) plots the results with and without the consideration of  ${J}^\mathrm{B}$.    As expected, the frequency of the eigen mode increases with $\Delta\varepsilon$ (thus, the current density).  An interesting point to note is that the addition of ${J}^\mathrm{B}$ appears to induce a comparatively larger jump in the frequency with a steeper slope than the corresponding change in the total current density.  For instance, the oscillation frequency at $\Delta\varepsilon = 30$ meV  goes up by about 40 \% (from 0.58 THz to 0.8 THz), while ${J}^\mathrm {B}$ only adds 13 \% to the total driving current.  The results clearly indicate that the bulk current, although only at a fraction in the magnitude of the surface term, can make a considerable contribution to the spin wave generation, highlighting the efficient nature of the SOT induced by the spin-Hall effect in the AFM/TI heterostructure.  With both the surface and bulk contributions, the oscillation frequency can reach 1 THz at the current density of $\sim 6\times10^7$~A/cm$^2$, which is nearly an order of magnitude smaller than the value estimated with a heavy metal as the SOC material.~\cite{Xu2019}   Once $\Delta\varepsilon$ becomes sufficiently negative, the spin wave excitation is no longer possible.  The corresponding threshold current density ($\sim 10^7$~A/cm$^2$)  is associated with the surface induced SOT via the anomalous Hall effect.

%%%%%%%%%% Oscillator
Along with electrical generation of traveling spin waves, spin torque nano-oscillators are another spintronic application that can take advantage of the AFM/TI system.  In this case, the AFM does not take a long strip form; rather the entire magnetic material is interfaced with and thus subject to excitation by the TI.   Unlike the structure described in Fig.~1, a hard-axis anisotropy is assumed in the $y$-axis ($K_y = -160$ kJ/m$^3$) normal to the directions of the driving current ($x$) and the spin current ($z$).  This hard-axis configuration is known to enable a low-threshold condition for oscillations as the N\'{e}el vector does not encounter the anisotropy energy barrier to rotate on the easy $x$-$z$ plane.  In addition, $\alpha$ is considered a variable while the rest of the parameters such as $A_\mathrm{ex}$ and  $M_s$ remain unchanged.  The N\'{e}el vector is assumed to be initially aligned in the $+x$ direction.

Figure~4 summarizes the multiplicity  of the N\'{e}el vector dynamics in the $\alpha$-$\Delta \varepsilon$ (thus, $J$) parameter space.  Two phase maps [Figs.~4(a) and 4(b)] correspond to the cases of TI electrons flowing in the $\mp x$ directions (thus,  $\hat{\mathbf{\sigma}} $ of $ \pm \hat{\mathbf{ y}} $), respectively.  Region 1 shows the conditions where the induced torque is insufficient to overcome the damping, resulting in no appreciable change in the magnetic state.  Once the induced SOT becomes sufficiently large to overcome the damping and other energy barriers, the N\'{e}el vector is driven from its initial orientation and rotates in the easy ${x}$-$ {z}$ plane (i.e., auto-oscillation; region 2). A further increase in the driving current can lead to a stable 90$^\circ$ rotation, aligning it along the injected spin polarization despite the hard-axis anisotropy in the same direction (region 3).  In the case of Fig.~4(b) with  $\hat{\mathbf{\sigma}} = - \hat{\mathbf{ y}} $, an additional dynamical behavior is observed along with the auto-oscillations and 90$^\circ$ rotations.  More specifically, 180$^\circ$ reversal of the N\'{e}el vector ($+ x \rightarrow -x$) can be realized in region 4.  The corresponding schematics are given in Figs.~4(c), 4(d), 4(e), and 4(f), respectively.

At first glance, it is not intuitively obvious for the system to have a directional preference in the $x$-$z$ plane enabling the stable rotation to $-x$ since no magnetic anisotropy is specified other than the hard $y$-axis.  However, the axial symmetry is broken by the explicit dependence of the surface-state anti-damping torque on the $z$ component of the magnetization unlike the SOT induced by a heavy metal (and thus the TI bulk states).  Since this term $\mathbf{T}_\mathrm{ad}^\mathrm{S}$ reduces to zero as the magnetization at the interface orients normal to the $z$ axis [i.e., $m_z \approx 0$; see Eq.~(\ref{Tsur})], the $\pm x$ directions serve in effect as the easy axis, offering bistable configurations.  In comparison, this reversal from $+x$ to $-x$ is not observed in Fig.~4(a) since the torque is induced toward the $+x$ axis (i.e., the initial orientation), which is the opposite direction to that experience in Fig.~4(b).  As such, the orientation remains unchange until strong excitation sufficiently disrupts the state leading to the auto-oscillation.  When the N\'{e}el vector is initially in the $-x$ direction, the dependence on the polarity of the driving current or electron flow is also reversed.  In such a set-up, the equivalent of Fig.~4(a) shows region 4 while that of Fig.~4(b) does not.  These results clearly suggest the possibility to deterministically encode the N\'{e}el vector orientation:  In the $\alpha$-$\Delta \varepsilon$ parameter space corresponding to region 4, the final state always aligns with the direction of the driving current irrespective of the initial orientation ($\pm x$).~\cite{Semenov2014}  Note that uncontrollable flip-flop's occur near the boundary with region 2 (i.e., just before the auto-oscillations) in both Figs.~4(a) and 4(b). As it involves a rather narrow range, this feature is not shown in the phase maps in order not to complicate the picture.  Similarly, no rapid change takes place across the boundary between regions 2 and 3. The steady oscillations do not disappear suddenly. Instead, the precession angle around the $\pm y$ axis gradually collapses from 90$^\circ$ (i.e., the $x$-$z$ plane) to nearly 0$^\circ$.  The oscillation frequency is clearly a strong function of $\alpha$ and $\Delta \varepsilon$ (i.e., the driving current density) with a wide tunable range.
The desired frequencies near the THz appear to require $\alpha$ smaller than those plotted ($ < 0.005$).  Along with the dynamic modulation of $\Delta \varepsilon$, the numerical value for $\alpha$  can also be tailored through doping or by introducing auxiliary layers.~\cite{Luo2014,Sun2013}

\section{Summary}
The feasibility of spin wave generation via the SOT induced in an adjacent TI is theoretically demonstrated in an easy-axis AFM. The results from the numerical simulations clearly show that the spin auto-oscillations can be achieved in the thin-film AFM strip through localized excitation, traveling over a long distance as the angle of precession gradually collapses due to the non-zero damping.  The calculations also elucidate the dependence of the N\'{e}el vector dynamics on the relevant physical parameters including the TI electronic properties, highlighting the potential significance of the bulk states as the source of the SOT in the heavily degenerate conditions.  Along with the propagating spin waves, the application of the AFM/TI bilayer structure as a spin torque nano-oscillator is also illustrated.   With the contributions from both the strongly spin-orbit coupled surface and bulk states, the TIs offer a highly efficient alternative to the conventional heavy metals for the SOT.

\begin{acknowledgments}
This work was supported, in part, by the US Army Research Office (W911NF-16-1-0472).
\end{acknowledgments}

\vspace{20pt}
\noindent
\textbf{DATA AVAILABILITY}

The data that support the findings of this study are available within the article.

\clearpage

\begin{figure}
\includegraphics[width=8cm]{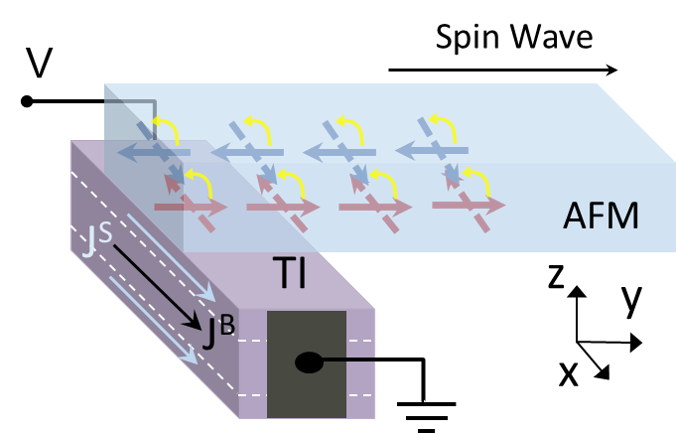}
\caption{Schematic illustration of the AFM/TI heterostructure under consideration (not to scale). The driving current flows along the ${x}$ axis in the TI (via the surface $ {{J}^\mathrm {S}}$ and bulk ${{J}^\mathrm {B}}$), inducing the spin current and the corresponding SOT in the AFM thin film with easy-axis ($y$) anisotropy.  The N\'{e}el vectors in the region of the AFM driven by the SOT undergo rotations and form spin waves that propagate along the strip.  The use of an A-type AFM provides ferromagnetic intra-plane coupling and, thus, net non-zero magnetization at the interface with the TI.}
\end{figure}

\clearpage

\begin{figure}
\includegraphics[width=8cm]{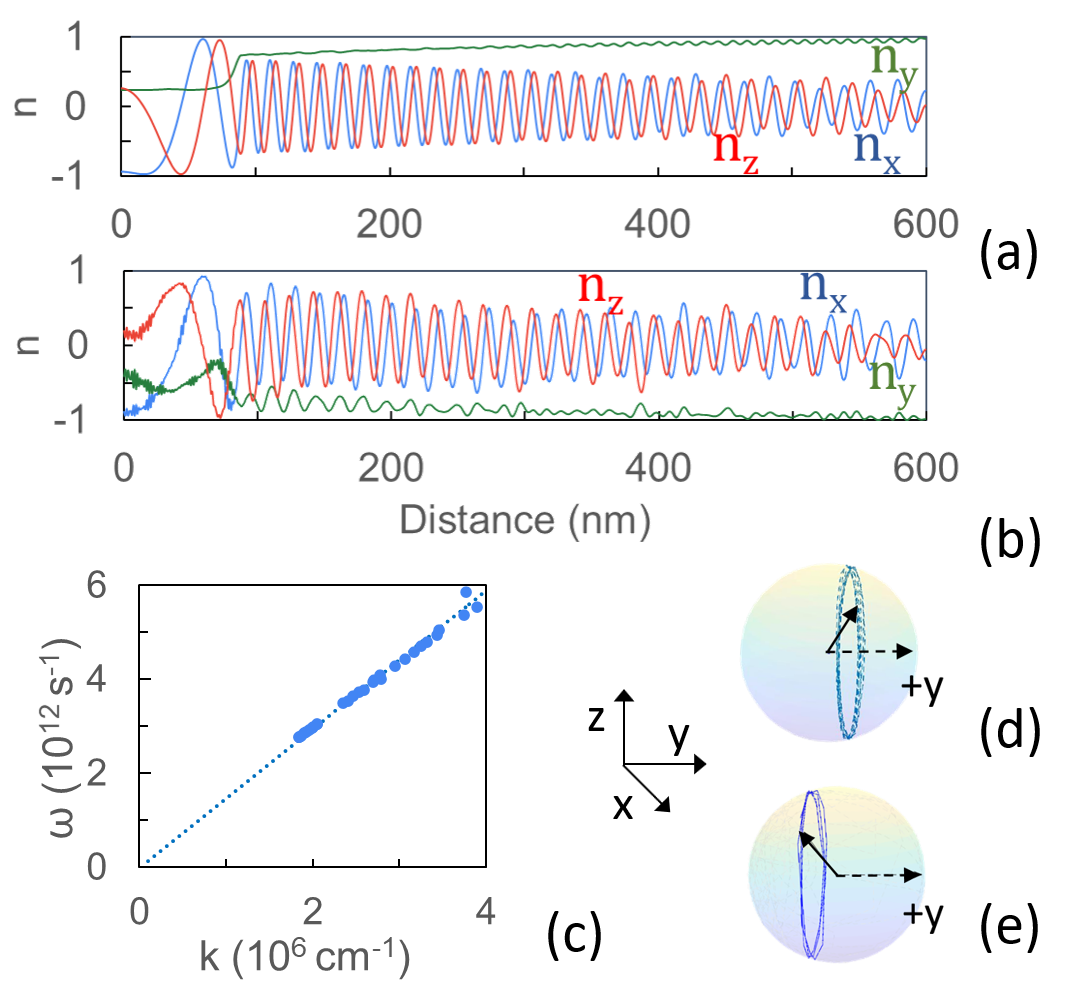}
\caption{(a,b) Snap shots of the steady-state N\'{e}el vector obtained along the AFM waveguide for two different directions of the spin current polarization $\hat{\mathbf{\sigma}} $ ($\pm {y}$).  With $\varepsilon_F - \varepsilon_C =30$ meV, the excitation current density (magnitude) corresponds to ${J}^\mathrm{B}=5.3\times10^6$ A/cm$^2$ and  ${J}^\mathrm{S}=4.1\times 10^7$ A/cm$^2$.   (c) Calculated spin wave dispersion relation. (d,e) Trajectories of the N\'{e}el vector in the excitation region with $\hat{\mathbf{\sigma}} = \pm \hat{\mathbf{y}}$. The initial state is aligned along the $+y$ easy-axis.  In (e) [as well as in (b)], the N\'{e}el vector reversal to $-y$ is observed before the precession. }
\end{figure}

\clearpage

\begin{figure}
\includegraphics[width=8cm]{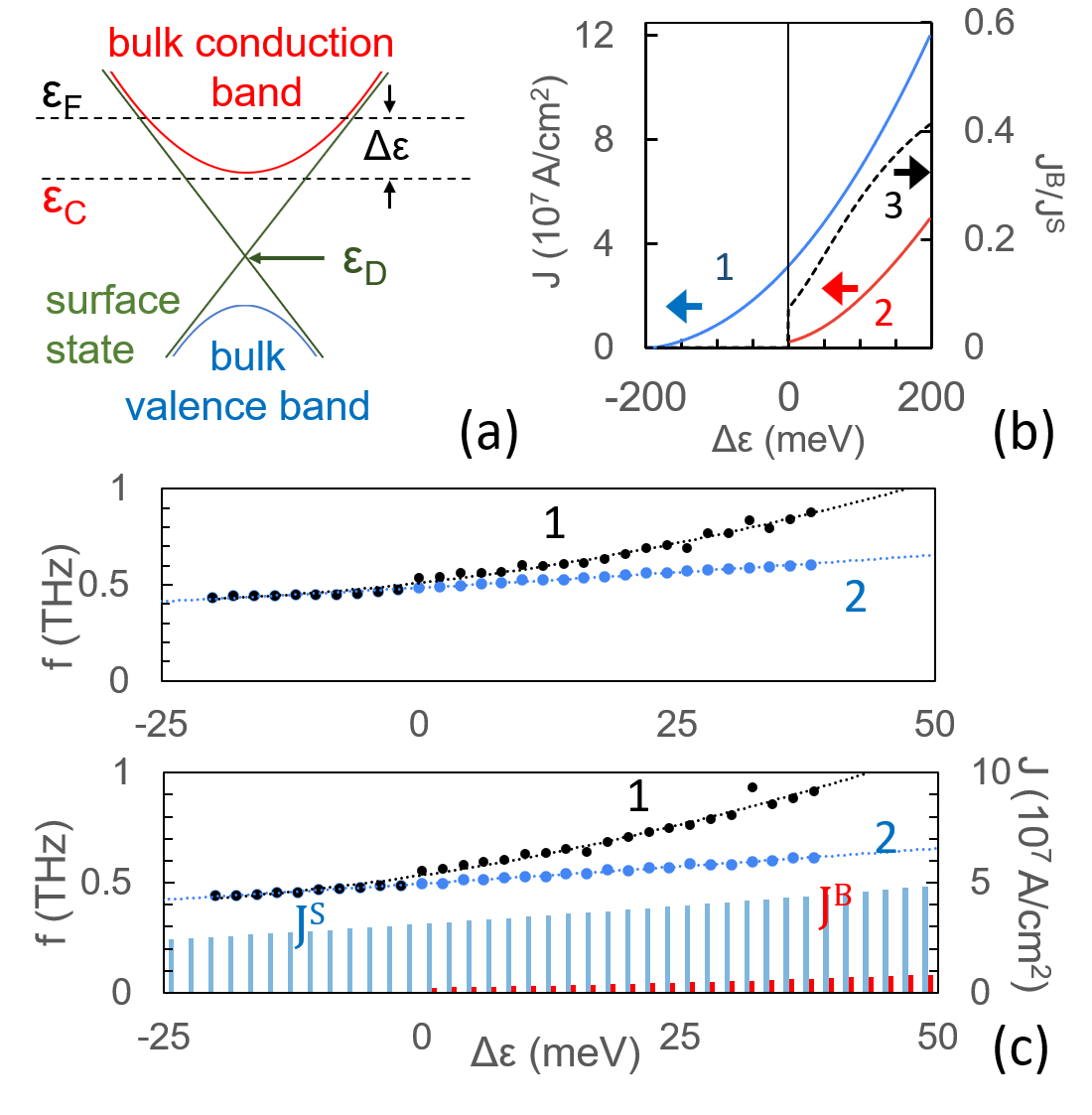}
\caption{(a) Typical electronic band structure of a TI.  The bulk states are approximated by parabolic energy bands, while the surface states are described by a linear dispersion with the Fermi velocity $v_F$.   $\Delta\varepsilon$ indicates the difference between the Fermi level $ \varepsilon_F$ and the bulk conduction band minimum $\varepsilon_C$. (b) Current density in the TI as a function of the Fermi level position. Line 1:  surface current density $J^\mathrm{S}$ in a 3D equivalent form, Line 2: bulk current density $J^\mathrm{B}$, and Line 3:  ratio of ${J}^\mathrm{B}/{J}^\mathrm{S}$.  (c) Excited spin wave frequency in the AFM as functions of the Fermi level position $\Delta\varepsilon$ with and without the SOT contributed by the bulk current (lines 1 and 2, respectively).  For convenience, the current density value corresponding to each  $\Delta\varepsilon$ is also provided for both $J^\mathrm{S}$ (blue) and $J^\mathrm{B}$ (red). The upper and lower panels correspond to the cases of  $\hat{\mathbf{\sigma}} = + \hat{\mathbf{y}}$  and $- \hat{\mathbf{y}}$, respectively. }
\end{figure}

\clearpage

\begin{figure}
\includegraphics[width=8cm]{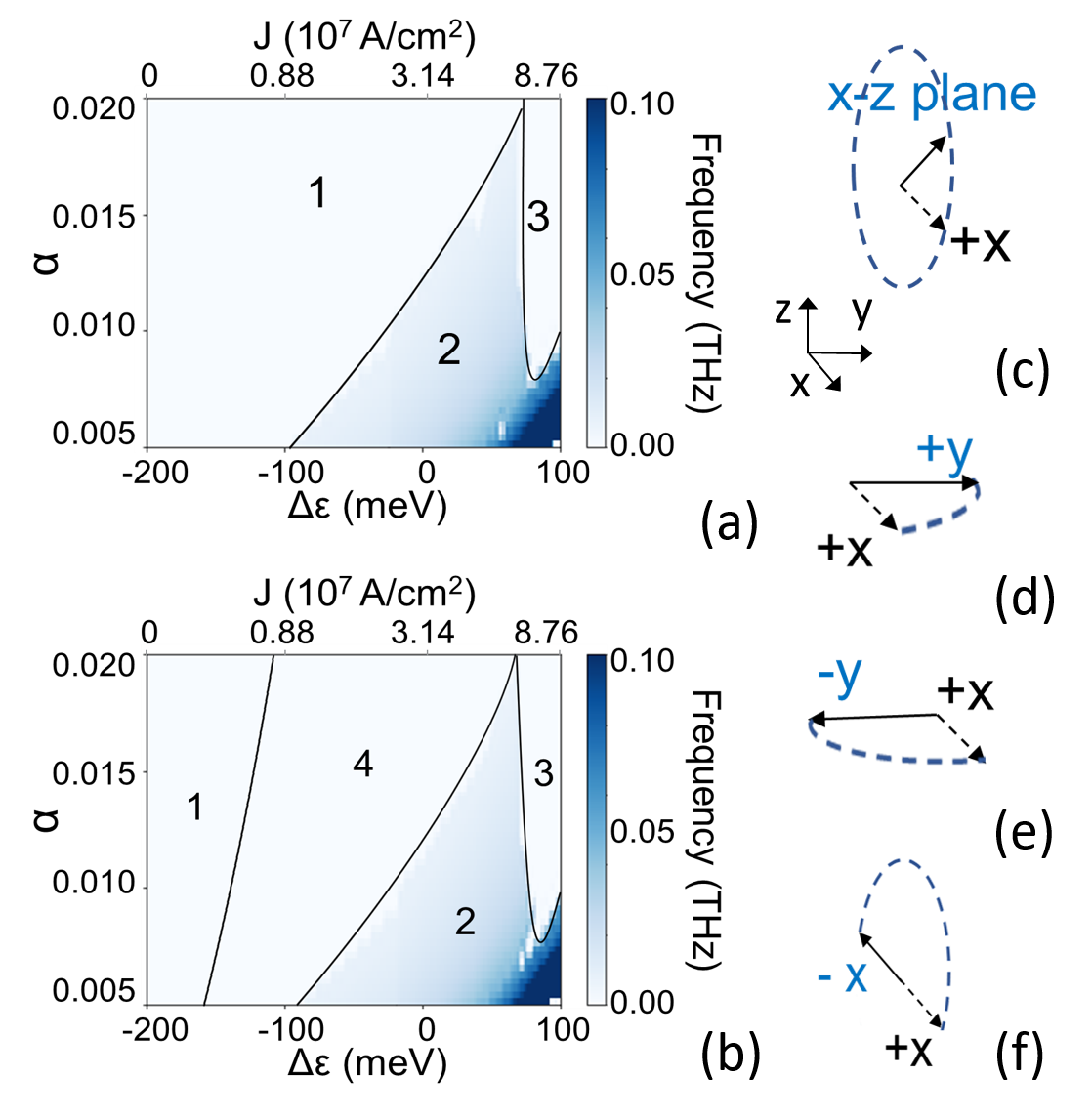}
\caption{N\'{e}el vector dynamics in the AFM/TI spin torque nano-oscillator.  A hard-axis anisotropy is assumed for the AFM in the $y$-axis ($K_y = -160$ kJ/m$^3$) normal to the directions of the driving current ($x$) and the spin current ($z$).  The entire magnet is subject to excitation by the SOT.   (a,b) Phase maps for various regions of operation in the $\alpha$-$\Delta \varepsilon$  parameter space with the spin current polarization  $\hat{\mathbf{\sigma}} = + \hat{\mathbf{y}}$  and $- \hat{\mathbf{y}}$, respectively. Regions 1, 2, and 3 correspond to the conditions of no appreciable change in the magnetic state, auto-oscillation, and 90$^\circ$ switching toward $\hat{\mathbf{\sigma}}$.  In the case of (b) with  $\hat{\mathbf{\sigma}} = - \hat{\mathbf{ y}} $, 180$^\circ$ rotation of the N\'{e}el vector ($+ x \rightarrow -x$) is observed before the auto-oscillations (region 4).  (c-f) Schematic illustration of the N\'{e}el vector motions characteristic to regions 2-4.}
\end{figure}

\end{document}